\documentclass[preprint,12pt,authoryear]{elsarticle}

\usepackage{graphicx}
\usepackage{natbib}
\usepackage{epstopdf}
\AppendGraphicsExtensions{.tif}

\usepackage{amsmath,amssymb}
\usepackage{floatrow}
\usepackage{subfig}

\usepackage{lineno}

\def\CO2{CO$_2$}
\newcommand{\perm}{k}

\makeatletter
\def\ps@pprintTitle{%
  \let\@oddhead\@empty
  \let\@evenhead\@empty
  \let\@oddfoot\@empty
  \let\@evenfoot\@oddfoot
}
\makeatother

\journal{Advances in Water Resources}

\begin{document}

\begin{frontmatter}



\title{Convective carbon dioxide dissolution in a closed porous medium at high-pressure real-gas conditions}

\author[label1,label2]{Baole Wen\corref{cor1}}
\ead{baolew@umich.edu}
\author[label3]{Zhuofan Shi} 
\author[label3]{Kristian Jessen} 
\author[label2,label4]{Marc A. Hesse}
\author[label3]{Theodore T. Tsotsis} 

\affiliation[label1]{organization={Department of Mathematics, University of Michigan},
            addressline={530 Church Street}, 
            city={Ann Arbor},
            postcode={48109}, 
            state={MI},
            country={USA}}        
\affiliation[label2]{organization={Department of Geological Sciences, Jackson School of Geosciences, The University of Texas at Austin},
            addressline={2275 Speedway Stop C9000}, 
            city={Austin},
            postcode={78712}, 
            state={TX},
            country={USA}}
\affiliation[label3]{organization={Mork Family Department of Chemical Engineering and Material Science, University of Southern California},
            addressline={925 Bloom Walk, HED 216}, 
            city={Los Angeles},
            postcode={90089}, 
            state={CA},
            country={USA}}    
\affiliation[label4]{organization={Oden Institute for Computational Engineering and Sciencess, The University of Texas at Austin},
            addressline={201 E 24th Street}, 
            city={Austin},
            postcode={78712}, 
            state={TX},
            country={USA}}
            
   
\cortext[cor1]{Corresponding author.}
            

\begin{abstract}
We combine modeling and measurements to investigate the dynamics of convective carbon dioxide (CO$_2$) dissolution in a pressure-volume-temperature cell, extending a recent study by \emph{Wen et al.} (\emph{J. Fluid Mech.}, vol. 854, 2018, pp. 56--87) at low-pressure under ideal-gas conditions to high-pressure and real-gas conditions.   Pressure-dependent compressibility and solubility are included to model the evolution of CO$_2$ concentration in the gas phase and at the interface, respectively.  Simple ordinary-differential-equation models are developed to capture the mean behavior of the convecting system at large Rayleigh number and are then verified by using both numerical simulations and laboratory experiments.  The prefactor for the linear scaling of convective CO$_2$ dissolution is evaluated -- for the first time -- by using pressure-decay experiments in bead packs at reservoir conditions.  The results show that our models could quantitatively predict the process of the convective CO$_2$ dissolution in pressure-decay experiments.  Moreover, the results also reveal that for increasing gas pressure in closed systems, the negative feedback of the pressure drop -- resulting from the dissolution of CO$_2$ in the liquid -- is weakened due to the decrease of the solubility constant at real-gas conditions.  Our analysis provides a new direction for determination and validation of the convective dissolution flux of CO$_2$ in porous media systems.
\end{abstract}


\begin{highlights}
\item Convective CO$_2$ dissolution in closed porous media is modeled at reservoir conditions
\item The quasi-steady convective dissolution flux of CO$_2$ is measured by experiments
\item The models are validated by using both laboratory experiments and numerical simulations
\item The negative feedback of the pressure drop is weakened at real-gas conditions
\item The analysis provides a new direction determining the convective dissolution flux
\end{highlights}

\begin{keyword}
convective dissolution, geological CO$_2$ storage, porous media convection, solubility trapping


\end{keyword}

\end{frontmatter}


\section{Introduction}\label{sec:introduction}
Solubility trapping is one of major trapping mechanisms in geological carbon dioxide (CO$_2$) storage, which is one promising means of reducing the emission of greenhouse gases into the atmosphere \citep{Metz2005,Orr2009}. After CO$_2$ is injected into the deep geological storage sites, e.g., the reservoirs or saline aquifers, it forms a less-dense CO$_2$-rich vapor phase accumulating at the top of the denser brine and below the low-permeability cap rock.  In this two-layer configuration, CO$_2$ dissolves into the brine and forms a diffusive boundary layer beneath the CO$_2$-brine interface. Since the resulting solution is denser than the underlying resident brine, convection may set in under the influence of the gravity when the diffusive boundary layer is thick enough \citep{Ennis-King2005,Riaz2006,Hassanzadeh2006,Xu2006,Slim2013}. This process can greatly increase the dissolution flux, reduce the possible leakage, and contribute to secure long-term storage \citep{Neufeld2010,Sathaye2014}.

To explore the dissolution process of CO$_2$ into the brine/water, many experiments have been performed in analog-fluid systems \citep{Neufeld2010,Backhaus2011,Tsai2013,MacMinnJuanes2013,Wang2016,Suekane2017,Liang2018,Salibindla2018,Liyanage2019,Michel-Meyer2020}, where the miscible analogue fluids, e.g., Methanol-Ethylene-Glycol-water and Propylene-Glycol-water, are used to replace CO$_2$ and brine. In these systems, the convective transport can be visualized in Hele-Shaw cells or using X-ray CT Scanning.  Nevertheless, there exist many fundamental differences between the analog-fluid systems and the CO$_2$-brine system.  The analog fluids and water are fully miscible, the density of the mixtures exhibits non-monotonic characteristics as a function of the solute mass fraction, and the interface moves upward during the mixing \citep{Neufeld2010,Hidalgo2015, Raad2016,Liang2018}. However, CO$_2$ is partially soluble in water, the density of the mixture increases monotonically with concentration, and CO$_2$-brine interface is nearly fixed in a closed system \citep{Duan2003,Efika2016,Shi2018,Wen2018JFM}.  Moreover, the analog-fluid systems allow the experiments at normal laboratory conditions, while the experiments with the CO$_2$-brine system at reservoir conditions involve a high-pressure and high-temperature setup \citep{Bickle2007,Shi2018}.

The major experimental approach used to measure the CO$_2$ dissolution in brine at reservoir conditions is the `pressure-decay' method \citep{Tse1979, Renner1988, Riazi1996, Wang1996, Zhang2000, Sheikha2005, Farajzadeh2009, Moghaddam2012, Azin2013,Mojtaba2014,Raad2015a, Zhang2015, Shi2018, Mahmoodpour2019,Mahmoodpour2020}. This approach is originally designed to characterize the \emph{diffusive} mass transfer of CO$_2$ in brine. In this method, the pressure in the gas phase above the liquid phase, due to the dissolution of CO$_2$, is monitored and recorded in a pressure-volume-temperature (PVT) cell (figure~\ref{fig:Geometry}). The evolution of the amount of CO$_2$ present in the gas phase is then calculated from the pressure, temperature and the volume of the gas phase via an equation of state. The experimental data is then commonly interpreted via a \emph{diffusion} model (i.e., the Fick's law), and the CO$_2$ diffusivity in the liquid phase is evaluated by fitting the experimental data.  Nevertheless, when the density-driven convection is introduced in the dissolution process, molecular diffusion is not the only mechanism that controls the CO$_2$ mass transfer.  In a few studies, researchers interpret the experimental data based on the diffusion model and then use a so-called `effective diffusivity' to describe the combined effect of diffusion and convection \citep{Yang2006,Farajzadeh2009,Wang2013,Zhao2018}.  In strong convecting systems, however, the reported values of the effective diffusivity are typically two orders of magnitude larger than the expected CO$_2$ diffusivity in bulk water or brine in a purely diffusive system \citep{Shi2018}.  Moreover, unlike the analog-fluid experiments which are convenient to evaluate the mean dissolution flux in the `constant-flux' regime at certain large Rayleigh number, in pressure-decay experiments the driving force for convection always decreases with time due to the pressure drop, leading to decreasing Rayleigh number and dissolution flux.  Hence, there exists a clear need today for developing appropriate models and theories to analyze the mixing processes of convective CO$_2$ dissolution in pressure-decay experiments at reservoir conditions.

Although extensive numerical simulations have been performed by several research groups to explore the dynamics of convective CO$_2$ dissolution in water \citep{Ennis-King2005, Hassanzadeh2006,Riaz2006,Xu2006,Pau2010,Hewitt2013shutdown,Szulczewski2013,Slim2014,DePaoli2016,DePaoli2017}, in their models a constant CO$_2$ concentration is usually assumed at the top boundary (i.e., the CO$_2$-water interface). However, in both the pressure-decay experiments and some natural CO$_2$ reservoirs \citep{Sathaye2014,Sathaye2016a,Sathaye2016b,Akhbari2017}, the systems are closed, so that the CO$_2$ dissolution reduces the pressure in the vapor phase, subsequently leading to a decline of the CO$_2$ concentration at the interface and thereby reducing the density difference driving convective dissolution in the liquid phase.  \citet{Wen2018JFM} investigate this negative feedback of the pressure drop on convective dissolution in closed systems at low pressures  (below 1 MPa) and present mathematical models for the dissolution process when the CO$_2$ vapor can be treated as an ideal gas.  Nevertheless, most natural gas reservoirs and pressure-decay experiments exist/operate under high-pressure conditions where the CO$_2$ vapor is no longer ideal.

In this investigation, we extend the modeling of convective CO$_2$ dissolution in closed porous media systems to high-pressure and real-gas conditions. Pressure-dependent compressibility factor and solubility are added in the ideal-gas law and the Henry's law, respectively, to model the evolution of CO$_2$ concentration in the gas phase and at the gas-liquid interface. Scaling analysis reveals that the dynamics of the system is controlled by five dimensionless parameters in a wide domain.  Both direct numerical simulations (DNS) and pressure-decay experiments are utilized to characterize the convective transport of CO$_2$ dissolution in closed porous media systems. Moreover, simple ordinary-differential-equation (ODE) models are developed to capture the mean behavior of the convecting system at large Rayleigh number and are then verified by simulations and experiments.  The mathematical models presented here have practical significance for exploring the dynamics of convective CO$_2$ dissolution in both pressure-decay experiments and confined reservoirs.  Importantly, our analysis allows the recovery of the constant-flux regime due to a rescaling, so that the pressure-decay experiments can be used to determine and validate the convective CO$_2$ dissolution flux as a function of Rayleigh number in porous media systems.

The remainder of this paper is organized as follows.  In the next section, we formulate the mathematical model of convection in closed porous media systems at high-pressure and real-gas conditions, non-dimensionalize the governing equations, describe the numerical methods used for the simulations, and construct the ODE models to predict the evolution of the gas pressure and dissolution flux in time. In section~3, we discuss the set-up of the pressure-decay experiments in bead packs. The numerical and experimental results are shown and discussed in section~4. Our conclusions are given in section~5.

\section{Mathematical modeling}\label{sec:FORMULATION}

Numerical simulations by \citet{Pau2010} and \citet{Fu2013} indicate that solutal convection in two-dimensional (2D) porous media systems exhibits similar flow regimes as in the three-dimensional (3D) systems, although in the quasi-steady regime the \CO2 dissolution flux of the latter is approximately 18\%--25\% higher than that of the former.  In this section, we focus on the modeling of convective \CO2 dissolution in a 2D rectangular domain.

\subsection{Dimensional equations}\label{sec:Dimen_Eqns}
\begin{figure}[t]
    \center{\includegraphics[width=0.8\textwidth]{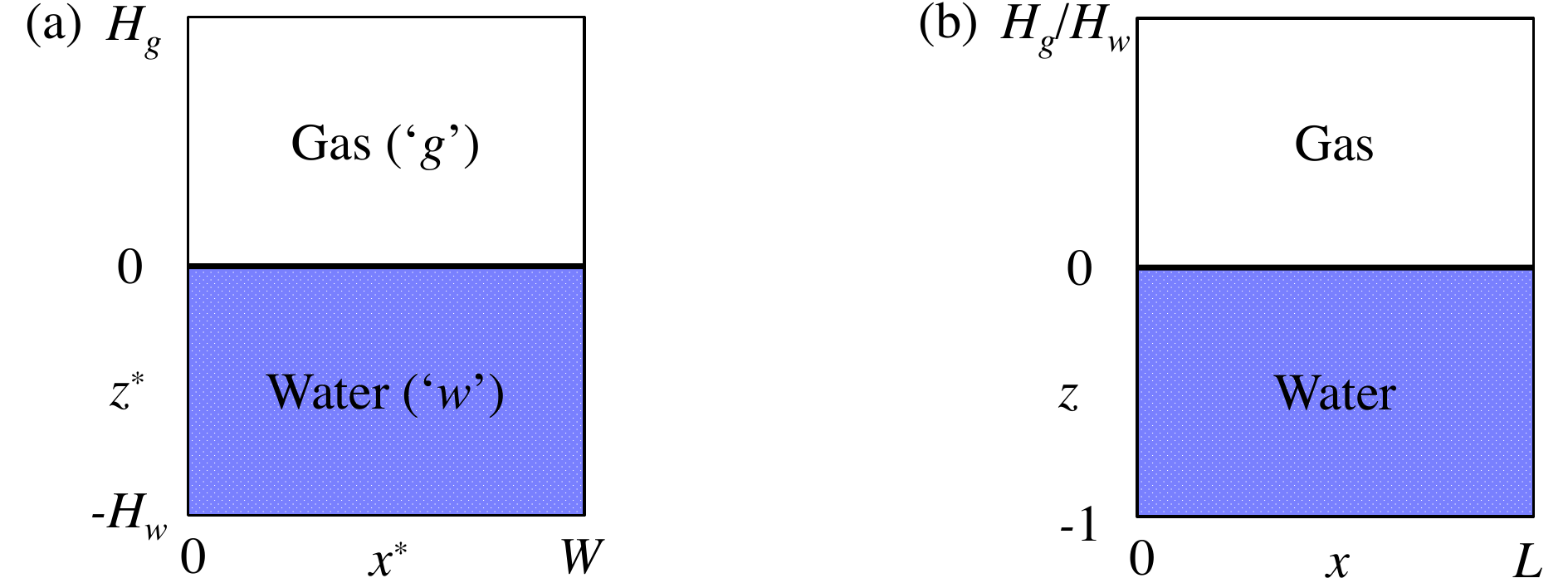}}
    \caption{Schematic of the two-dimensional pressure-decay cell. (a) The dimensional domain has heights $H_g$ and $H_w$ for the gas and water fields, respectively, and width $W$; (b) the dimensionless domain has an aspect ratio $L = W/H_w$ for the water field.}  \label{fig:Geometry}
\end{figure}

Consider a 2D isotropic and homogeneous porous medium containing gas overlying water (figure~\ref{fig:Geometry}a). We neglect the capillary forces so that the phases are segregated by buoyancy and separated by a sharp interface at $z_0^* = 0$ \citep{Golding2011,Martinez2016}. In the gas phase (i.e., $0 < z^* < H_g$), the CO$_2$ vapour is assumed to be a real gas,
\begin{eqnarray}
	P_g^*V_g^* = n_gZRT, \label{RealGas}
\end{eqnarray}
where $P_g^*$ is the gas pressure, $V_g^*$ is the gas volume, $n_g$ is the amount of gas in moles, $Z$ is the gas compressibility factor measuring how much the gas deviates from ideal-gas behavior \citep{RealGas2020}, $R$ is the universal gas constant, and $T$ is the absolute temperature. At the gas-water interface, the local equilibrium between the gas and the dissolved aqueous \CO2 is given by Henry's law
\begin{eqnarray}
	C_{s}^* =  K_h P_g^*, \label{Henry}
\end{eqnarray}
where $C_{s}^*$ is the dissolved gas concentration (or the \CO2 concentration at the interface) and $K_h$ is the Henry's law solubility constant.  In previous study at low pressure \citep{Wen2018JFM}, the gas is assumed to be ideal, so $Z\equiv~1$ and $K_h$ is constant (at isothermal conditions). For high-pressure real-gas conditions, however, both $Z$ and $K_h$ are pressure-dependent, i.e., $Z= Z(P_g^*)$ and $K_h= K_h(P_g^*)$.  Hence, equation~(\ref{Henry}) yields $C_s^* = K_h(P_g^*)P_g^*$.  Moreover, in the modeling of convection the volume change of water due to the \CO2 dissolution is negligible under certain pressure condition (approximately 3.5\% for fixed $P_g^*=12$ MPa), so the interface is assumed to be always localized at $z^*=0$ in closed systems \citep{Wen2018JFM}.  Nevertheless, the solubility $K_h$ used here is from previously published batch phase-equilibrium data which have already considered the volume change of water. 

In the liquid phase (i.e., $-H_w < z^* < 0$), the flow is incompressible and obeys Darcy's law
\begin{eqnarray}
	&\nabla^*\cdot\mathbf{u}_w^* = 0,\label{Continuity}\\
	&\mathbf{u}_w^* = - \dfrac{\perm}{\mu\phi} \left(\nabla^*{P}_w^* + \rho_w^*g{\bf e}_{z^*}\right),\label{Darcy}
\end{eqnarray}
where $\mathbf{u}_w^* = (u^*,w^*)$ is the volume-averaged pore velocity, $\perm$ is the medium permeability, $\mu$ is the dynamic viscosity of the fluid, $\phi$ is the porosity, $P_w^*$ is the pressure in the water field, $g$ is the gravity acceleration term and ${\bf e}_{z^*}$ is a unit vector in the $z^*$ direction.  The density of the \CO2-water solution, $\rho_w^*$, is assumed to be a linear function of the concentration
\begin{eqnarray}
	\rho_w^* =  \rho^*_0 + \Delta\rho^*_0\dfrac{C^*_w}{C_{s, 0}^*},\label{Rho}
\end{eqnarray}
where $\rho_0^*$ is density of the fresh water, $C_w^*$ is the concentration of dissolved \CO2 in the water, and $\Delta \rho^*_0$ and $C_{s, 0}^*$ are, respectively, the density difference between the fresh water and the saturated water and the concentration of \CO2 in saturated water at the initial gas pressure $P_{g,0}^*$.  In this work, $\Delta \rho^*_0$ is modeled using a quadratic function of $P_{g}^*$ described in \citet{Wen2018JFM}.  The transport of the concentration field in the liquid phase is governed by the following advection-diffusion equation
\begin{eqnarray}
\dfrac{\partial{C_w^*}}{\partial t^*} + \nabla^*\cdot\left(\mathbf{u}_w^*C_w^*\right) = D{\nabla^*}^2{C_w^*}, \label{Solute}
\end{eqnarray}
where $D$ is the diffusivity.  From \citet{Shi2018}, $D \approx 3.37 \times 10^{-9}$ m$^2$/s for $\phi\approx0.4$ at $T=323.15$ K (50$^\circ$C).

Initially, the water contains no dissolved gas,
\begin{eqnarray}
	\left.C_w^*\right|_{t^*=0}  = 0\;\; \mbox{for} \;\; -1 < z^* < 0. 
	\label{IC_pde}
\end{eqnarray}
For boundary conditions of the liquid phase, the upper boundary (i.e., the \CO2-water interface at $z^*=0$) is determined by local equilibrium with the gas and is impermeable to fluid; and the lower boundary is impermeable to solute and fluid:
\begin{eqnarray}
	 \left.C_w^*\right|_{z^* = 0} =  C_{s}^*(t^*), \quad \left.w^*\right|_{z^* = 0} = 0; \quad 
	 \left.\dfrac{\partial C^*_{w}}{\partial z^*}\right|_{z^* = -H_w} =0, \quad   \left.w^*\right|_{z^* = -H_w} = 0. \label{eq:BC}
\end{eqnarray}
All fields satisfy an $W$-periodicity condition in the $x^*$ direction. In the closed system, the mass balance for the gas and liquid phases requires 
\begin{eqnarray}
	\frac{d n_g}{d t^*} = -AF^*, \label{GasMole}
\end{eqnarray}
where the area of the interface $A=W$ in the 2D system, and the molar flux from the gas into the water $F^*$ can be evaluated as 
\begin{eqnarray}
	F^* =  \left.D\frac{\partial \overline{C_w^*}}{\partial z^*}\right|_{z^* = 0} = \frac{D}{W}\int_0^W\left.\frac{\partial C_w^*}{\partial z^*}\right|_{z^* = 0}dx^*, \label{Flux}
\end{eqnarray}
where `$\overline{\;\cdot\;}$' denotes the horizontal average as defined above.

\begin{figure}[t]
    \center{\includegraphics[width=0.95\textwidth]{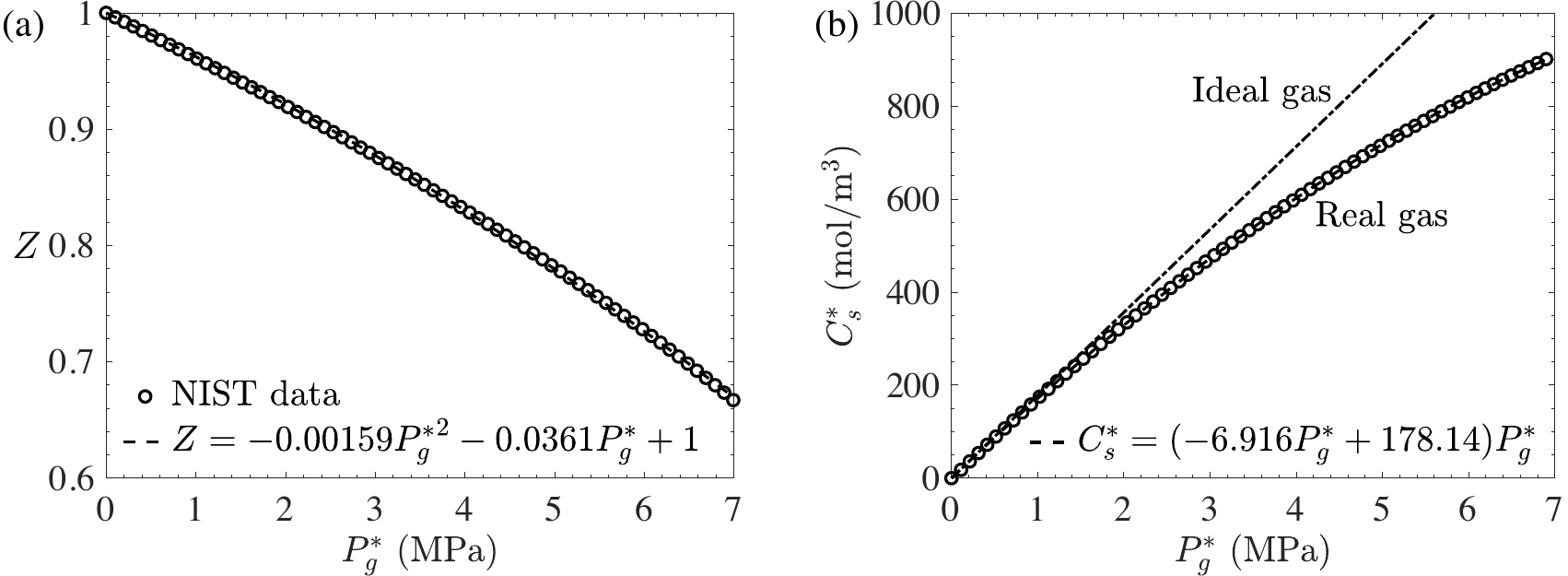}}
    \caption{Correlations for (a) compressibility factor $Z$ and (b) dissolved gas concentration $C_s^*$ as function of gas pressure $P_g^*$ at $T = 323.15$ K.}  \label{fig:Z_Cs}
\end{figure}

In this investigation, we model $Z$ and $C_s^*$ with correlations developed from previously published batch phase-equilibrium (NIST) data, namely, for fixed temperature $T$,
\begin{eqnarray}
   Z = z_1{P_g^*}^2 + z_2{P_g^*} +1 \quad \mbox{and} \quad C_s^* = K_hP_g^* = (k_1P_g^* + k_2)P_g^*, \label{Z_Cs_models}
\end{eqnarray}
where $z_1 = -0.00159$ MPa$^{-2}$, $z_2 = -0.0361$ MPa$^{-1}$, $k_1 = -6.916$ mol m$^{-3}$ MPa$^{-2}$ and $k_2 = 178.140$ mol m$^{-3}$ MPa$^{-1}$ at $T = 323.15$ K.  As shown in figure~\ref{fig:Z_Cs}, at low pressure (e.g., $P_g^* \lesssim 1$ MPa), $Z\approx 1$ and $C_s^*$ increases linearly with $P_g^*$, so the gas reduces to the ideal state.  At high pressure, however, the gas deviates from the ideal-gas behavior and $C_s^*$ levels off.


From equations (\ref{RealGas}) and  (\ref{GasMole})--(\ref{Z_Cs_models}), we can write the evolution of the gas pressure as
\begin{eqnarray}
   \dfrac{d P_g^*}{d t^*} = -\dfrac{(z_1{P_g^*}^2 + z_2{P_g^*} +1)^2 R T A D}{V_g^*}\dfrac{1}{1 - z_1{P_g^*}^2}\dfrac{\partial\overline{C_w^*}}{\partial {z^*}}|_{z^* = 0}. \label{Pg}
\end{eqnarray}

\subsection{Dimensionless equations}\label{sec:Dimenless_Eqns}
In the liquid phase, the system is rendered dimensionless using the water-layer thickness $H_w$ (figure~\ref{fig:Geometry}b), the buoyancy velocity $U = \perm\Delta\rho_0^* g/(\mu\phi)$, the diffusive time $\mathcal{T} = H_w^2/D$, the pressure $\mathcal{P} = \Delta\rho_0^*gH_w$, the density difference $\triangle \rho_0^*$ and the concentration of \CO2 in saturated water $C_{s, 0}^*$ at $P_{g,0}^*$; in the vapor phase, the pressure is normalized using the initial gas pressure $P_{g,0}^*$. Then, we have dimensionless variables 
\begin{eqnarray}
	\mathbf{x} = \dfrac{\mathbf{x}^*}{H_w},\;\mathbf{u} = \dfrac{\mathbf{u}^*}{U},\;t = \dfrac{t_w^*}{\mathcal{T}},\;\tilde{P} = \dfrac{P_w^*}{\mathcal{P}},\;\rho = \dfrac{\rho_w^*}{\triangle \rho_0^*},\;C = \dfrac{C_w^*}{C_{s, 0}^*},\;P_g = \dfrac{P_g^*}{P_{g,0}^*}. \label{scales}
\end{eqnarray}
Upon substitution, the flow and mass transport processes of the system are governed by the following dimensionless equations
\begin{eqnarray}
&\nabla\cdot\mathbf{u} = 0, \label{Continuity_nondim}\\
& \mathbf{u} = -{\nabla}P - C{\bf e}_{z}, \label{Darcy_nondim} \\
& \dfrac{\partial C}{\partial t} + {Ra_0}\mathbf{u}\cdot\nabla C = {\nabla}^2 C, \label{Solute_nondim}
\end{eqnarray}
where $P = \tilde{P} + (\rho_0^*/\Delta\rho_0^*)z$ and $Ra_0$ is the initial Rayleigh-Darcy number
\begin{eqnarray}
Ra_0 = \frac{\perm\Delta\rho_0^* g H_w}{\phi\mu D}.\label{eq:Ra0}
\end{eqnarray}
In dimensionless form, the initial condition and boundary conditions become  
\begin{eqnarray}
	C|_{t=0}  = 0\;\; \mbox{for} \;\; z < 0, 
	\label{ICs_nondim}
\end{eqnarray}
and 
\begin{eqnarray}
	\left.C\right|_{z=0} =  C_s(t), \quad \left.w\right|_{z=0} = 0; \quad \left.\dfrac{\partial C}{\partial z}\right|_{z=-1} = 0, \quad \left.w\right|_{z=-1} = 0,  \label{BCs_nondim1}
\end{eqnarray}
where the dimensionless dissolved concentration at the interface $C_s$ is given by 
\begin{eqnarray}
	C_s = \dfrac{K P_g^2 +  P_g}{K + 1},  \label{Cs_Pg}
\end{eqnarray}
with
\begin{eqnarray}
	K = \dfrac{k_1}{k_2} {P_{g,0}^*}  \label{K_para}
\end{eqnarray}
representing the correction of the Henry's law solubility constant at high pressure.  The evolution of the dimensionless gas pressure $P_g$ is then governed by the following ODE and initial condition 
\begin{eqnarray}
   \dfrac{d P_g}{d t} = -\Pi_0 \dfrac{(Z_1P_{g}^2 + Z_2P_{g} + 1)^2}{1 - Z_1P_{g}^2} F \quad \mbox{and} \quad \left.P_g\right|_{t=0} = 1, \label{Pg_nodim}
\end{eqnarray}
where 
\begin{eqnarray}
   \Pi_0 = \dfrac{V_w^*}{V_g^*}(k_1 P_{g,0}^* + k_2)RT \label{Pi_nodim}
\end{eqnarray}
is the `dissolution capacity' of the water volume $V_w^* = H_wW\phi$ in the 2D rectangular domain,
\begin{eqnarray}
   Z_1 = z_1 {P_{g,0}^*}^2 \quad \mbox{and} \quad Z_2 = z_2 P_{g,0}^* \label{Z1_Z2_nodim}
\end{eqnarray}
represent the correction of the compressibility factor at high pressure, and 
\begin{eqnarray}
    F = \left.\dfrac{\partial\overline{C}}{\partial {z}}\right|_{z = 0} \label{F_nondim}
\end{eqnarray}
is the dimensionless dissolution flux. Therefore, in an infinitely-wide domain ($L\rightarrow\infty$), the dynamics of the system are controlled by the dimensionless parameters $Ra_0$, $\Pi_0$, $Z_1$, $Z_2$, and $K$.  At $P_g^* \lesssim 1$ MPa, $Z_1$, $Z_2$ and $K$ are all approximately zero, so that the dimensionless system reduces to the low-pressure case presented by \citet{Wen2018JFM}.

\subsection{Numerical method}\label{sec:Numerical}
We perform direct numerical simulations for the dimensionless, partial-differential-equation (PDE) model (\ref{Continuity_nondim})--(\ref{F_nondim}) by adapting the solver developed in previous investigations~\citep{WenChini2018JFM,Shi2018,Wen2018JFM,Wen2018PRFluids}. For the 2D system, we introduce the stream function $\psi$ to describe the fluid velocity, i.e., $\mathbf{u} = (u,w) = (\partial_z\psi, -\partial_x\psi)$, so that the continuity equation (\ref{Continuity_nondim}) is satisfied a priori. Here $\psi$ is $L$-periodic in $x$ and homogeneous at the top and bottom of the water layer.  The resulting partial differential equations (\ref{Darcy_nondim}) and (\ref{Solute_nondim}) are discretized in space using a Fourier--Chebyshev-tau spectral algorithm \citep{Boyd2000}, and discretized in time using a third-order-accurate semi-implicit Runge--Kutta scheme for the first three steps \citep{Nikitin2006} and a fourth-order-accurate semi-implicit Adams--Bashforth/backward-differentiation scheme for the remaining steps \citep{Peyret2002}.  $P_g$ and $C_s$ are updated at each time step by solving equations~(\ref{Cs_Pg}) and (\ref{Pg_nodim}) using a two-step Adams--Bashforth algorithm.

\subsection{ODE models}\label{sec:ODE}
At high Rayleigh numbers, solutal convection in porous media exhibits different flow regimes characterized by the behavior of the dissolution: An early diffusion-dominated regime, a flux-growth \& plume-merging regime, a quasi-steady convective regime, and a shut-down regime~\citep{Riaz2006,Hewitt2013shutdown,Tilton2014,Slim2014,Wen2018JFM}.  At early time, the transport is dominated by diffusion and the flux decays as $(\pi t)^{-1/2}$.  As the diffusive boundary layer near the interface becomes sufficiently thick, convective instability sets in: Small finger plumes are generated, leading to an increase of the flux. After a series of plume-merging events, the resulting primary plumes propagate at a constant speed when the gas pressure remains constant ($\Pi_0=0$), resulting in a constant flux (i.e., the quasi-steady convective regime). After the primary plumes reach the bottom of the water layer, the \CO2-rich fluid starts to move upwards with the returning flow. Once this dense fluid reaches the top boundary (i.e., the interface), there is no fresh water in the liquid phase, so that the density difference (i.e., the driving force) for convection decreases, the flux declines, and eventually the convection shuts down.  

In \citet{Wen2018JFM}, simple ODE models are developed to predict the \CO2 dissolution in closed porous media systems at low-pressure, ideal-gas conditions for both quasi-steady convective and shut-down regimes. These models are based on certain properties of the convective flow at large $Ra_0$ and are extended to the high-pressure real-gas conditions in this manuscript. 

In the quasi-steady convective regime, the analysis and simulation results in \citet{Wen2018JFM} show that the dissolution flux is proportional to the square of the interface concentration, more specifically
\begin{eqnarray}
   F = \alpha Ra_0 C_s^2, \label{F_Cs2}
\end{eqnarray}
where the prefactor $\alpha\approx 0.017$ from previous investigations \citep{Hesse2008thesis,Pau2010,Hewitt2013shutdown,Slim2014,Wen2018JFM} with constant \CO2 concentration at the top boundary (i.e., $C_s=1$). Equation~(\ref{F_Cs2}) will be verified later via laboratory experiments at high-pressure conditions. Actually, the pressure conditions of the vapor phase only affect the equation of state~(\ref{RealGas}) and the solute solubility at the interface~(\ref{Henry}), while the behavior described by equation~(\ref{F_Cs2}) is an inherent characteristic for the high-Rayleigh-number convective flow (in the liquid phase) and, therefore, can be directly utilized for the modeling at high-pressure conditions.  Combining equations~(\ref{Cs_Pg})--(\ref{F_nondim}) with equation~(\ref{F_Cs2}) yields an ODE model for $P_g$ in the quasi-steady convective regime:
\begin{eqnarray}
\dfrac{d P_g}{d t}  = - \alpha Ra_0\Pi_0 \dfrac{(Z_1 P_{g}^2 + Z_2 P_{g} + 1)^2}{1 - Z_1 P_{g}^2} \left(\dfrac{KP_g^2 +  P_g}{K + 1}\right)^2 \;\; \mbox{and} \;\; \left.P_g\right|_{t=0} = 1.
\label{Pg_ODE_quasisteady}
\end{eqnarray}
Once $P_g$ is known, $C_s$ and $F$ can be evaluated from equations (\ref{Cs_Pg}) and (\ref{F_Cs2}).

When the convection is shut-down, numerical simulations by \citet{Hewitt2013shutdown} and \citet{Wen2018JFM} indicate the horizontal-mean \CO2 concentration $\overline{C}$ becomes vertically well-mixed, so that 
\begin{eqnarray}
   \overline{C}(t) \approx \overline{\overline{C}}(t) \equiv \dfrac{\mathbf{\int} Cd\mathbf{x}}{V_w}, \label{C_volavg}
\end{eqnarray}
where `$\overline{\overline{\;\cdot\;}}$' denotes the volume-averaged concentration in water as defined above and $V_w$ is the normalized water volume.  Hence, we can write 
\begin{eqnarray}
	F = \dfrac{d}{dt} \Large{\int}_{-1}^0\overline{C}dz = \dfrac{d\overline{\overline{C}}}{dt}. \label{Flux_nondim2}
\end{eqnarray}
Following \citet{Hewitt2013shutdown} and \citet{Wen2018JFM}, we define a time-dependent Nusselt number
\begin{eqnarray}
	Nu(t) \equiv \dfrac{F}{C_s - \overline{\overline{C}}} = \beta Ra_e = \beta (C_s - \overline{\overline{C}})Ra_0, \label{Nu}
\end{eqnarray}
where $Nu$ varies as a linear function of the effective Rayleigh number $Ra_e$ as defined above and $\beta$ is an undetermined constant.  Since the mass conservation of the whole system requires
\begin{eqnarray}
	\frac{P_g}{Z(P_g^*)} + \Pi_0\overline{\overline{C}}  = \frac{1}{Z(P_{g,0}^*)}, \label{Pg_Cvolavg}
\end{eqnarray}
combining equations~(\ref{Cs_Pg}) and (\ref{Flux_nondim2})--(\ref{Pg_Cvolavg}) yields an ODE model for $P_g$ in the shut-down regime:
\begin{eqnarray}
\dfrac{d P_g}{d t}  =  - \beta Ra_0\Pi_0 \dfrac{(Z_1 P_{g}^2 + Z_2 P_{g} + 1)^2}{1 - Z_1 P_{g}^2} \left[\dfrac{K P_g^2 +  P_g}{K + 1} + \;\dfrac{1}{\Pi_0}\dfrac{P_g}{Z_1 P_{g}^2 + Z_2 P_{g} + 1} \right. \nonumber\\
\left. - \dfrac{1}{\Pi_0}\dfrac{1}{Z_1 + Z_2 + 1}\right]^2.
\label{Pg_ODE_shutdown}
\end{eqnarray}
For consistency of the ODE models at high- and low-pressure conditions, we set $\beta = 0.0317$, and then solve equations~(\ref{Cs_Pg}) and (\ref{Pg_ODE_shutdown}) numerically subject to the following initial condition:
\begin{eqnarray}
C_{s,0} = \dfrac{1}{1+\Pi_0}\left[1 + \dfrac{\Pi_0}{c_0}\right],
\end{eqnarray}
where $c_0 = 0.861$ \citep{Wen2018JFM}.
With $P_g$ and $C_s$, $\overline{\overline{C}}$ and $F$ can be determined by solving equations (\ref{Nu}) and (\ref{Pg_Cvolavg}).

\section{Experimental set-up}\label{sec:Exps}
\begin{figure}[t]
    \center{\includegraphics[width=1\textwidth]{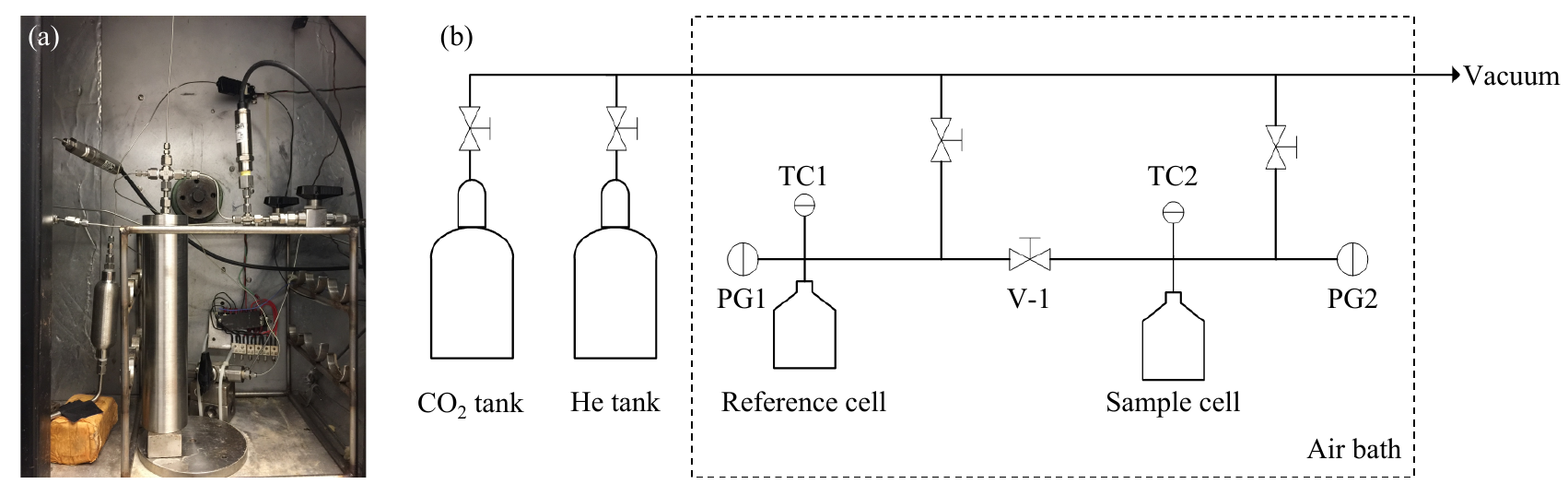}}
    \caption{Experimental set-up. (a) Experimental apparatus; (b) schematic diagram.}  \label{fig:Exp_setup}
\end{figure}

The experiments were carried out in a typical PVT system that mainly consists of a reference cell and a sample cell (figure~\ref{fig:Exp_setup}), with further details having been described previously \citep{Shi2018}. A known mass of water was first loaded into the sample cell, its volume then calculated from its known density. Subsequently, a known mass of soda-lime glass beads with a particle size of 1.6 mm was added to the sample cell (wet packing) until the height of the beads equalled the water level. The height of the water-saturated porous material was then recorded and the (bulk) volume of the porous material was calculated from the cell geometry. The porosity of the unconsolidated porous medium $\phi$ was then calculated from the volume of the water $V_w^*$ and the bulk volume of the porous material, and is reported in table~\ref{table:Exp_para}. The gas volume for each experiment was then calculated from the known volume of the sample cell and the total volume of the porous material. The permeability of the porous medium $k$ is estimated using the Kozeny--Carman correlation \citep{Wyllie1955}
\begin{eqnarray}
\perm = \dfrac{d^2\phi^3}{180(1-\phi)^2},
\end{eqnarray}
where $d$ is the bead diameter.

\begin{table}
\centering
\begin{tabular}{l c c c c c c c}
\hline
Case & $H_w$ (cm) & $A$ (cm$^2$) & $\phi$     & $P_{g,0}^*$ (MPa) & $V_w^*/V_g^*$ & $\Pi_0$ & $Ra_0$  \\
\hline
  Exp. 1        & 5.667  & 6.421  & 0.406    & 5.553     & 0.670    & 0.26   & 18169 \\
  Exp. 2        & 12.034 & 6.379  & 0.385    & 2.159     & 2.135    & 0.97   & 15349 \\
  Exp. 3        & 24.016 & 6.379  & 0.374    & 2.509     & 4.538    & 2.02   & 32102 \\
  Exp. 4        & 23.962 & 6.379  & 0.367    & 2.434     & 6.085    & 2.72   & 29145 \\
\hline
\caption{Information about the experiments.  Exp.~1 was conducted by \citet{Shi2018}.}
\label{table:Exp_para}
\end{tabular}
\end{table}

All the experiments were performed at a constant temperature of 323.15~K. To initiate each experiment, the reference cell was pressurized with \CO2 to a certain pressure and allowed to reach thermal equilibrium. The valve in between the reference cell and the sample cell was then opened to allow the \CO2 in the reference cell to enter the sample cell. The valve was then closed when the pressure in the sample cell reached the initial pressure shown in table~\ref{table:Exp_para} (in a few seconds), so that the reference cell was isolated from the sample cell for the duration of the experiment. From that point on, the pressure of \CO2 in the gas space in the sample cell started to decrease due to its dissolution into the water-saturated porous medium. For the remainder of the experiment the gas pressure $P_g^*$  and the temperature $T$ in the sample cell were then monitored and recorded continuously.

We report here data from four experiments each employing a different height of porous medium.  The information of each experiment is listed in table~\ref{table:Exp_para}.  The first experiment (Exp.~1) was conducted previously by \citet{Shi2018}, while the remaining three experiments (Exp.~2 to Exp.~4) were conducted specifically for this study (a different cell was used for Exp.~1 with a slightly larger cell cross-sectional area).  In Exp.~2--Exp.~4, Teflon spacers were used to reduce the volume of the sample cell in order to arrive at a different water-gas volume ratio ($V_w^*/V_g^*$) for each experiment.  

\section{Results and discussion}\label{sec:Results}

\begin{figure}[t]
    \center{\includegraphics[width=0.95\textwidth]{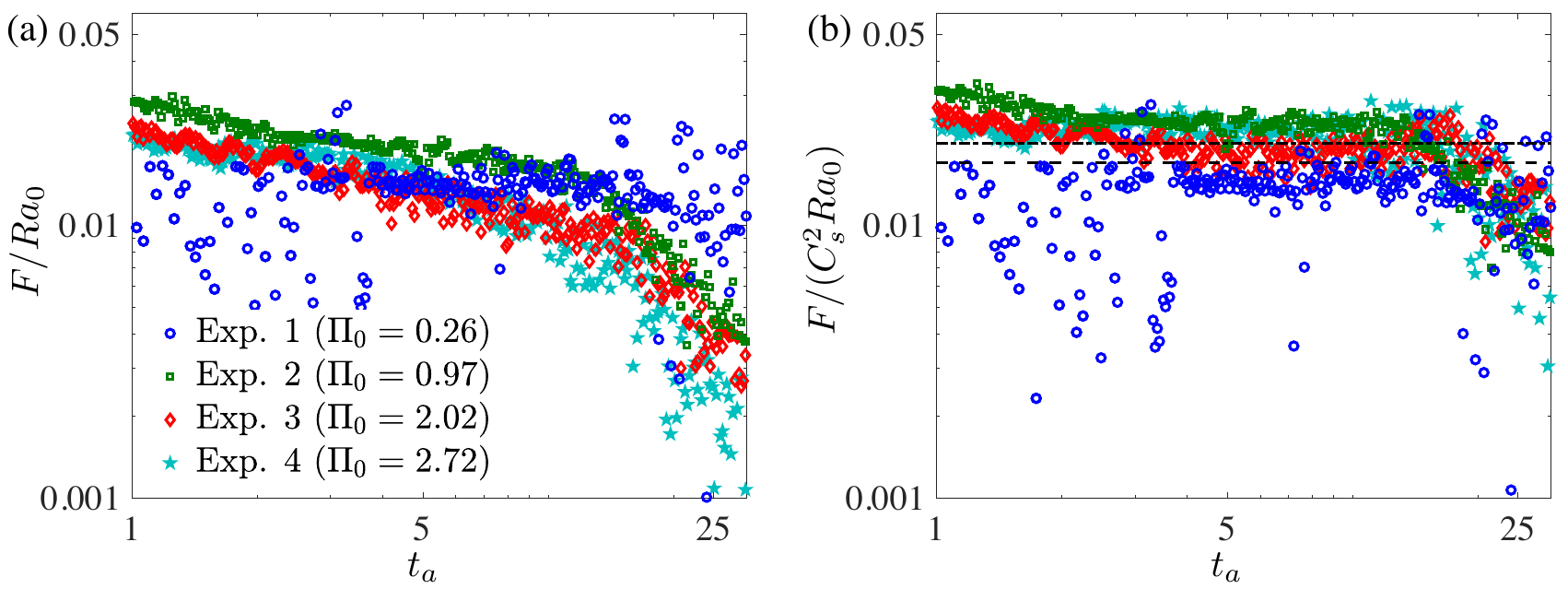}}
    \caption{Evolution of the compensated dissolution flux in time from experiments. (a): In closed systems, the dissolution flux declines with time due to the pressure drop in the vapor phase; (b) under appropriate rescaling, the constant-flux characteristic, i.e., $\alpha~=~F/(C_s^2Ra_0)$ shown in equation~(\ref{F_Cs2}), is recovered in the quasi-steady convective regime ($1 \lesssim t_a \lesssim 20$). From Exp.~1 to Exp.~4, $\alpha \approx 0.012$, 0.023, 0.019 and 0.022, respectively.   Dash line: $\alpha = 0.017$ from 2D DNS \citep{Pau2010,Hewitt2013shutdown,Slim2014,Wen2018JFM}; dash-dot line: $\alpha = 0.02$ from 3D DNS \citep{Fu2013}.}  \label{fig:Flux_tc_EXPs}
\end{figure}

Here we use experiments to first quantitatively verify the high-$Ra_0$ property of the dissolution flux during convection as proposed in equation~(\ref{F_Cs2}).  Figure~\ref{fig:Flux_tc_EXPs} shows the evolution of the flux $F$ compensated by $Ra_0$ and $C_s^2Ra_0$ as a function of the advection time $t_a=Ra_0 t$,  as observed from the experiments.  In the experiments, the moles of \CO2 in the vapor phase are determined based on the pressure and temperature data ($\sim40000$ points for Exp.~1 and $\sim120000$ points for each of Exp.~2--Exp.~4) via equations~(\ref{RealGas}) and (\ref{Z_Cs_models}).  The dimensional flux  $F^*$ is then evaluated from the time derivative of the amount of \CO2 in the vapor phase via equation~(\ref{GasMole}), to arrive at the dimensionless flux $F = H_wF^*/(DC_{s,0}^*)$.  In figure~\ref{fig:Flux_tc_EXPs}, the scatter in the flux is due to a simple finite-difference approximation of the derivative.  

In closed systems, the dissolution of \CO2 into the underlying water reduces the gas pressure.  The pressure drop in the gas phase decreases the saturated concentration of \CO2 in the water, and introduces a negative feedback that reduces the convective dissolution flux.  As shown in figure~\ref{fig:Flux_tc_EXPs}(a), for increasing dissolution capacity $\Pi_0$ the decline of the dissolution flux is more evident, especially at the late time.  After appropriate rescaling, however, $F/(C_s^2Ra_0)$ becomes nearly constant in the quasi-steady convective regime ($1 \lesssim t_a \lesssim 20$), consistent with the theoretical analysis which finds that $F/C_s^2 = \alpha Ra_0$.  Our four experiments reveal that the prefactor $\alpha$ varies from 0.012 to 0.023, in close agreement with the 2D \& 3D DNS predictions $\alpha=0.017$ and 0.02 for a constant \CO2 concentration at the top boundary (see figure~\ref{fig:Flux_tc_EXPs}b).  To the best of our knowledge, this is the first experimental study to evaluate the prefactor of $F\sim \alpha Ra_0$ for high-Raleigh-number convective \CO2 dissolution at reservoir conditions.

In previous experiments, the convective dissolution flux of \CO2 in porous media systems was usually measured using analog fluids in Hele-Shaw cells or granular porous media \citep{Neufeld2010,Backhaus2011,Liang2018}.  The analog-fluid system can successfully visualize and characterize the convective mixing process at atmospheric-pressure conditions.  Nevertheless, the nonlinear density dependence of miscible analog fluids induces a wavy, upward-moving interface and produces a higher prefactor $\alpha$ for the dissolution flux than that in the real \CO2-water system.  Based on our analysis and validation above, the pressure-decay experiments can be used to determine the convective \CO2 dissolution flux in porous media systems at reservoir conditions.  

To further interpret the experiments, we perform high-resolution numerical simulations with both the PDE and ODE models based on the information listed in table~\ref{table:Exp_para}.  In order to avoid the scatter in the experimental flux, as seen in figure~\ref{fig:Flux_tc_EXPs}, in the rest of this section we fit the following smooth function to the experimental pressure data $P_g^*$ and moles of \CO2 in the vapor phase $n_g$,
\begin{eqnarray}
f = a_0 + a_1 e^{-a_2t^*} + a_3 e^{-a_4t^*} + a_5 e^{-a_6t^*},\label{fit_fun}
\end{eqnarray}
where $a_0$--$a_6$ are all positive numbers. Then, the dissolution flux can be evaluated by calculating the time derivative of equation~(\ref{fit_fun}) as discussed above.

\begin{figure}[t!]
    \center{\includegraphics[width=0.95\textwidth]{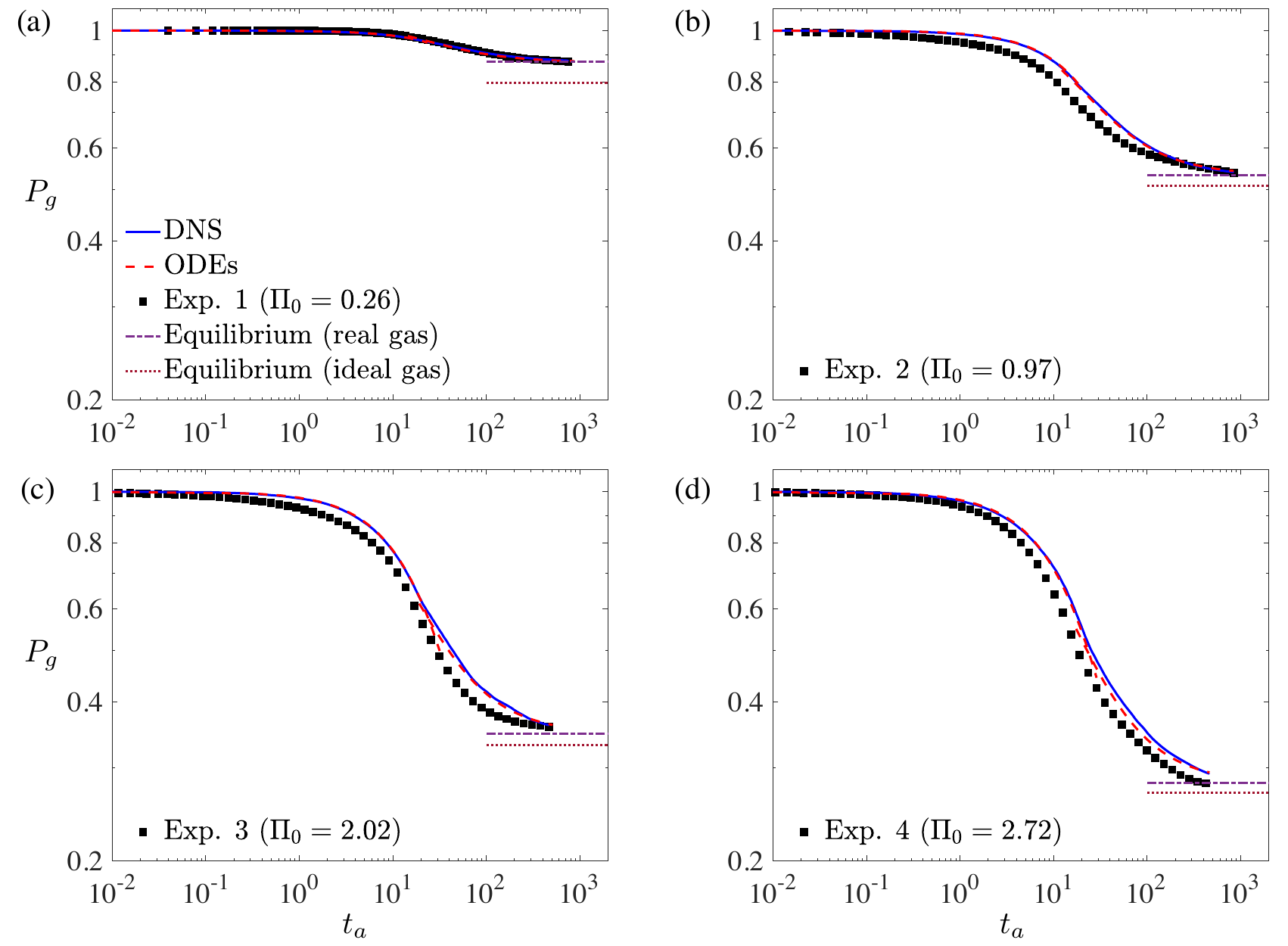}}
    \caption{Comparison of gas pressure $P_g$ from experiments, DNS and ODE models. Dash-dot and dot lines represent, respectively, the equilibrium pressure by solving the steady version of the real-gas model~(\ref{Pg_ODE_shutdown}) and from the ideal-gas model given by \citet{Wen2018JFM}, i.e., $P_g|_{t_a\rightarrow\infty}=1/(1+\Pi_0)$.}  \label{fig:Pgnondim_vs_tc_DNSvsEXPvsODE}
\end{figure}
\begin{figure}[h!]
    \center{\includegraphics[width=0.95\textwidth]{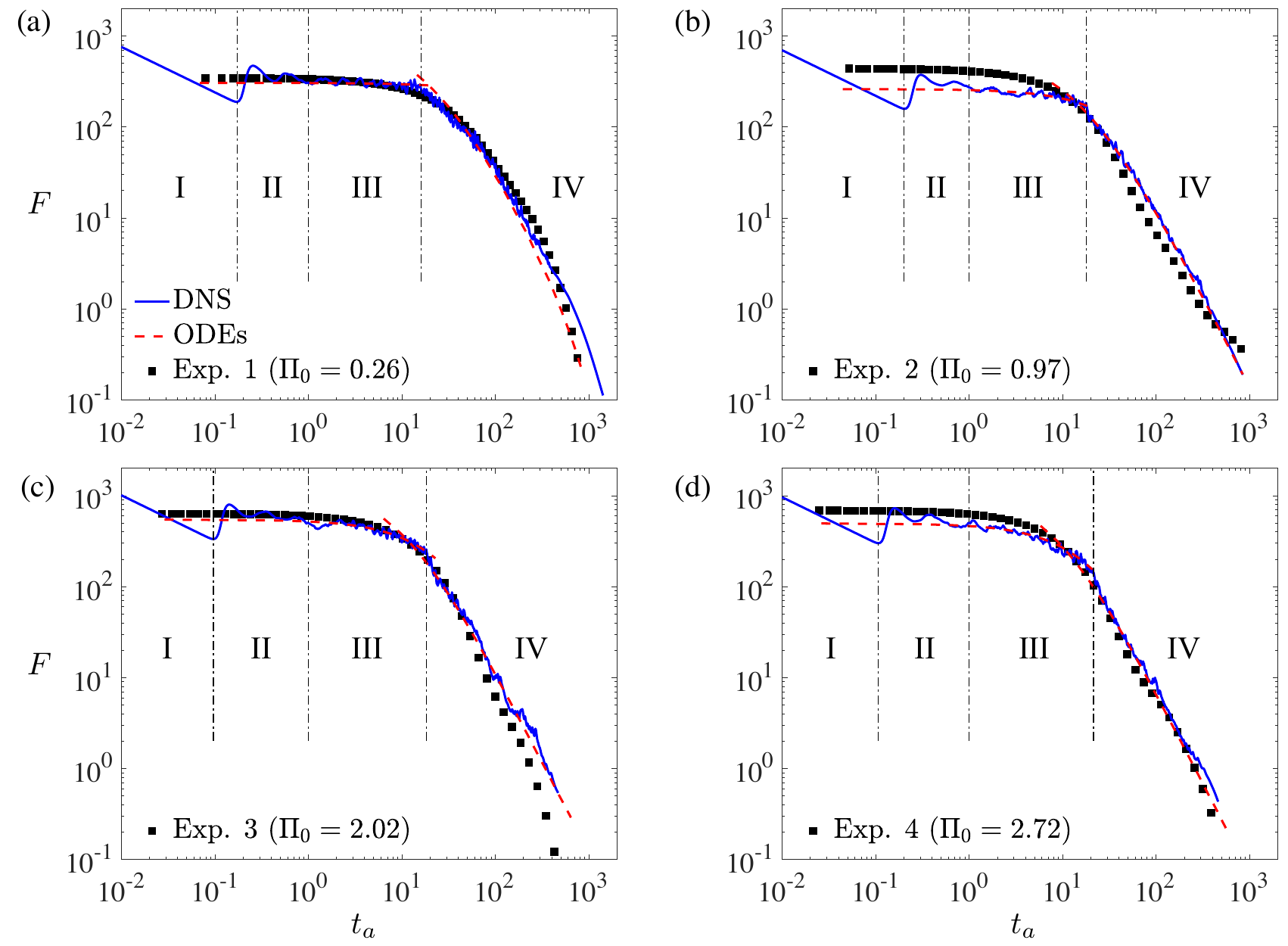}}
    \caption{Comparison of dissolution flux $F$ from experiments, DNS and ODE models. The four
dynamic regimes are delineated using dash-dotted lines based on the flow characteristics: (I) diffusion-dominated regime ($t_a\lesssim 3000/Ra_0$); (II) flux-growth \& plume-merging regime ($3000/Ra_0 \lesssim t_a\lesssim 1$); (III) quasi-steady convective regime ($1 \lesssim t_a\lesssim t_s$); and (IV) shut-down regime ($t_a\gtrsim t_s$), where $t_s$ is the onset of the shut-down regime from DNS. For (a)--(d), $t_s \approx$ 16.0, 17.9, 18.3 and 21.4, respectively.}  \label{fig:Fluxnondim_vs_tc_DNSvsEXPvsODE}
\end{figure}

Figures~\ref{fig:Pgnondim_vs_tc_DNSvsEXPvsODE} and \ref{fig:Fluxnondim_vs_tc_DNSvsEXPvsODE} compare the dimensionless gas pressure $P_g$ and dissolution flux $F$ for DNS of the PDE model, ODE models in quasi-steady and shut-down regimes, and the experimental results. As shown in these figures, the ODE models agree well with the DNS, while both models are found to be in good agreement with the experiments.  

For different values of $\Pi_0$, the gas pressure in figure~\ref{fig:Pgnondim_vs_tc_DNSvsEXPvsODE} has a similar evolution with time, but declines to lower values at higher $\Pi_0$.  In the log-log plot, the inflection point on the time-pressure curve corresponds to the transition from the quasi-steady convective regime to the shut-down regime.  Moreover, the steady state of equation~(\ref{Pg_ODE_shutdown}) successfully predicts the equilibrium gas pressure in the experiments.  Since the Henry's law solubility constant $K_h(P_{g}^*)$ for \CO2 in water decreases with increasing $P_{g}^*$ at real-gas conditions (figure~\ref{fig:Z_Cs}b), the equilibrium gas pressure from the real-gas model is always higher than that from the ideal-gas model, where $K_h$ assumes a constant value \citep{Wen2018JFM}.

\begin{figure}[t]
    \center{\includegraphics[width=0.95\textwidth]{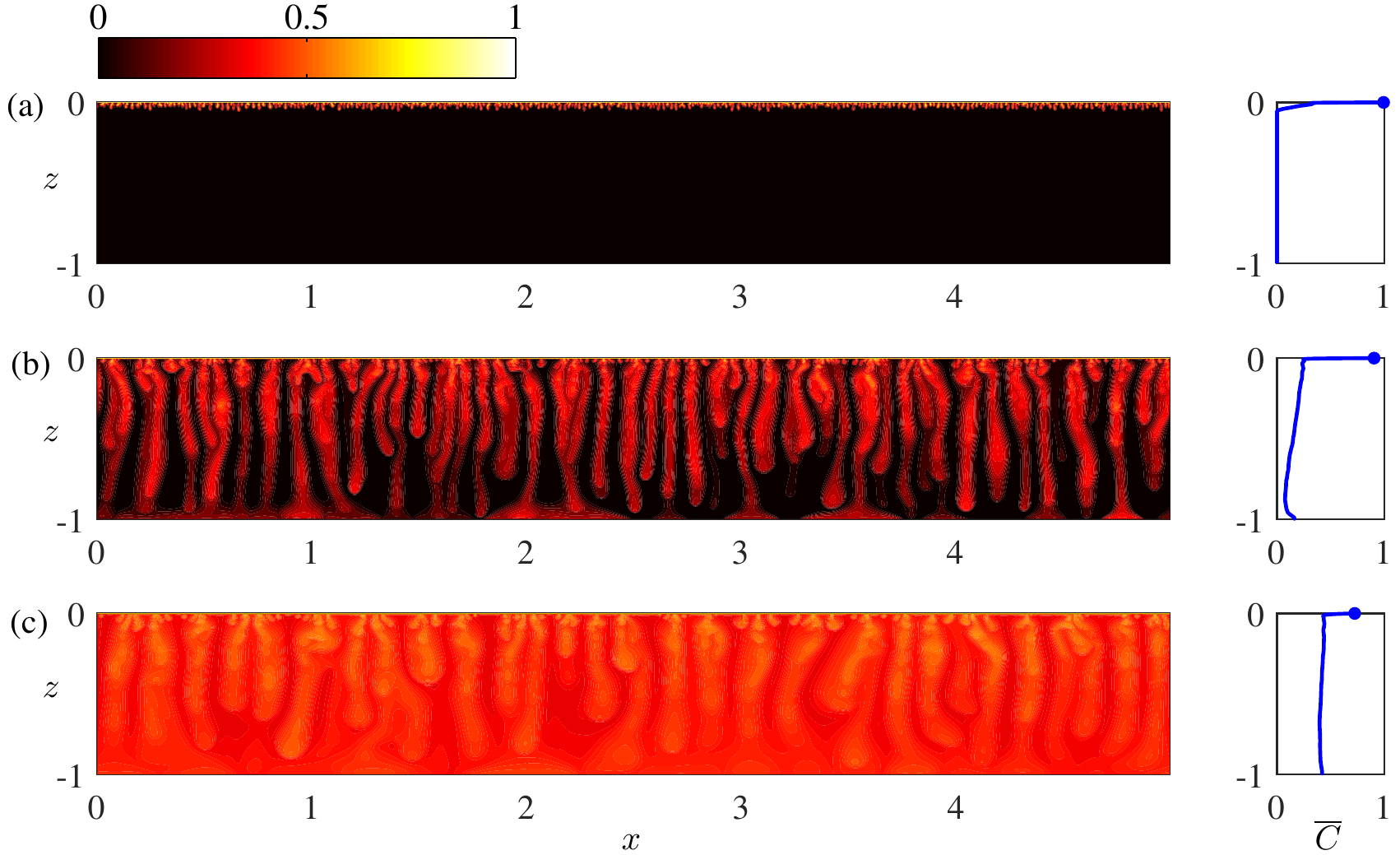}}
    \caption{Snapshots of the concentration field $C$ and the corresponding horizontal-mean concentration profile $\overline{C}$ from DNS at $Ra_0=20000$ with $\Pi_0=1$ and $P_{g,0}^* =$ 3.546 MPa: (a) $t_a=0.5$ (regime II); (b) $t_a=10$ (regime III); (c) $t_a=50$ (regime IV). The dot at the top of $\overline{C}$ denotes the concentration at the \CO2-water interface.}  \label{fig:C_Pi35atm_DNS}
\end{figure}

Consistent with the ideal-gas case \citep{Wen2018JFM}, four flow regimes are delineated based on the behavior of the dissolution as shown in figure~\ref{fig:Fluxnondim_vs_tc_DNSvsEXPvsODE}. (I) Diffusion-dominated regime: At early time (right after opening the valve), the dissolution is dominated by diffusion, so that the flux $F\sim(\pi t)^{-1/2}$; (II) flux-growth \& plume-merging regime: After a certain time, convection sets in due to the buoyancy-driven instability and small plumes are generated beneath the \CO2-water interface, growing laterally and merging with neighboring plumes (figure~\ref{fig:C_Pi35atm_DNS}a); (III) quasi-steady convective regime: After the convective pattern has coarsened, the basic pattern of the primary plumes is nearly unchanged (figure~\ref{fig:C_Pi35atm_DNS}b); and (IV) shut-down regime: Once the saturated fluid that is carried upward by the return flow reaches the interface, the driving force for convection (i.e., $\Delta C$) decreases, the flux declines rapidly, and eventually the convection shuts down (figure~\ref{fig:C_Pi35atm_DNS}c).  Despite the name of the convective regime (III), convection in closed systems is never actually quasi-steady as the flux declines due to the negative feedback of the pressure drop.  Moreover, the DNS results in figure~\ref{fig:Fluxnondim_vs_tc_DNSvsEXPvsODE} reveal that increasing $\Pi_0$ delays the transition to the shut-down regime.

To systematically investigate the dynamics of convective \CO2 dissolution in closed systems at high-pressure real-gas conditions, we perform DNS in the 2D rectangular domain at $Ra_0 = 20000$ and $L=5$ with $\Pi_0=0.5$, 1, 2 \& 5 and $P_{g,0}^* =$ 0.507 MPa (5~atm), 2.027 MPa (20~atm), 3.546 MPa (35~atm) \& 5.066 MPa (50~atm).  Figure~\ref{fig:C_Pi35atm_DNS} shows the snapshots of the concentration field $C$ and its horizontal-mean profile $\overline{C}$ at flow regimes (II)--(IV) with $\Pi_0=1$ and $P_{g,0}^*=3.546$ MPa.  The convective pattern at high-pressure real-gas conditions remains similar for various flow regimes as in the ideal-gas case discussed in \citet{Wen2018JFM}.  The concentration at the \CO2-water interface decreases with time due to the pressure drop in the vapor phase.  This reduces the driving force for convection, i.e., the concentration difference $\Delta C$ of the fluids between the interface and the underlying water, and thereby induces a negative feedback on the dissolution flux.  Moreover, as shown in figure~\ref{fig:C_Pi35atm_DNS}(c), in the shut-down regime the horizontal-mean concentration becomes well-mixed and nearly constant with depth, supporting the hypothesis used in building the ODE model in section~\ref{sec:ODE}.  We observe similar features for other values of $\Pi_0$ and $P_{g,0}^*$, which are omitted in this manuscript.

\begin{figure}[t]
    \center{\includegraphics[width=0.95\textwidth]{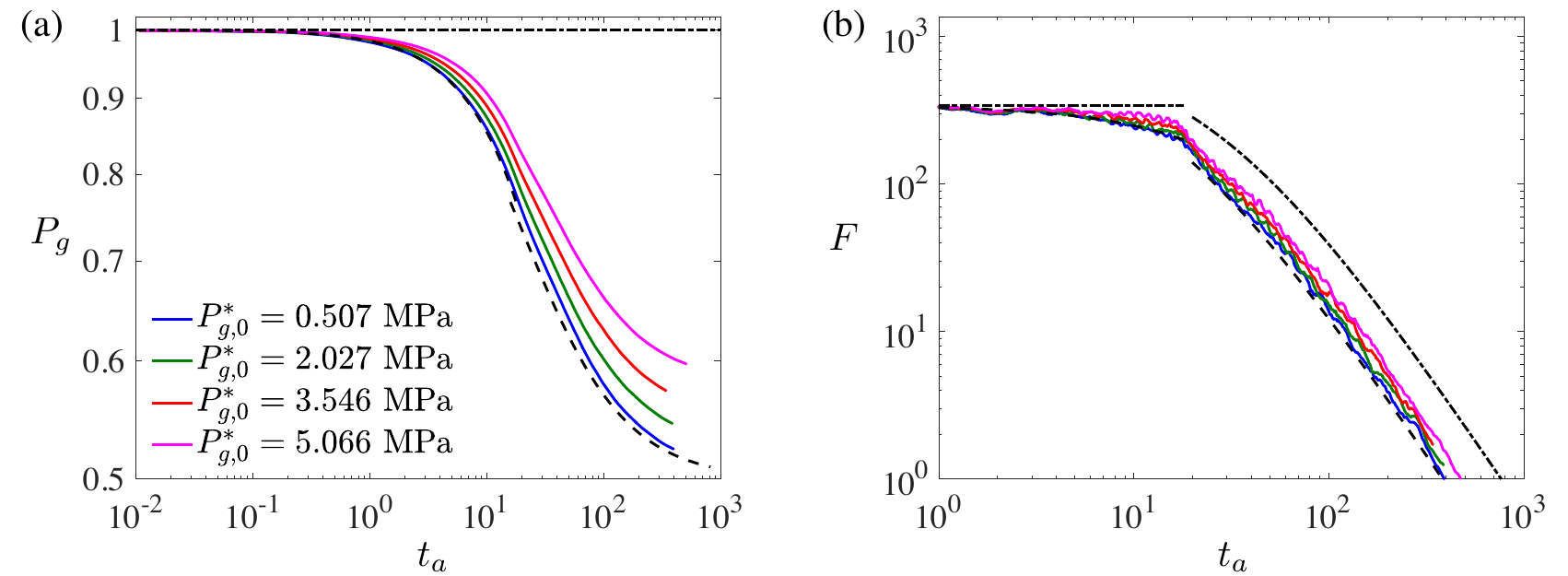}}
    \caption{Evolution of (a) gas pressure $P_g$ and (b) dissolution flux $F$ in time from numerical simulations at $Ra_0 = 20000$ and $\Pi_0=1$ with various $P_{g,0}^*$. Dashed line: ideal-gas ODE models with $\Pi_0=1$; dash-dot line: ODE models with $\Pi_0=0$ \citep{Wen2018JFM}.  \label{fig:PgF_vs_tc_DNSPi1}}
\end{figure}

Figure~\ref{fig:PgF_vs_tc_DNSPi1} shows the real-gas effects on the evolution of the gas pressure and dissolution flux at $Ra_0 = 20000$ and $\Pi_0=1$.  For reference, the ODE models for the ideal-gas case and the constant-pressure ($\Pi_0=0$) case are also plotted.  At low-pressure condition, e.g., $P_{g,0}^*=0.507$ MPa, the gas is ideal, so $P_g$ (which is equal to $C_s$) and $F$ are predicted accurately by the models from \citet{Wen2018JFM}.  As the pressure is increased, the gas deviates from the ideal state and the solubility constant $K_h(P_{g}^*)$ declines (figure~\ref{fig:Z_Cs}).  This reduced the pressure drop and the associated negative feedback on the dissolution flux; on the other hand, at $\Pi_0=0$ the pressure in the vapor phase remains constant and the negative feedback disappears.  Therefore, for closed systems at real-gas conditions, the ideal-gas models by~\citet{Wen2018JFM} provide lower bounds on the gas pressure and the dissolution flux, while the ODE models with $\Pi_0=0$ provide the corresponding upper bounds.

\section{Conclusions}\label{sec:Conclusion}
In this work, we extend the modeling of convective \CO2 dissolution in closed porous media systems to high-pressure real-gas conditions.  This is critical for the determination of convective CO$_2$ dissolution flux at reservoir conditions from pressure-decay experiments in a standard PVT cell.  Based on previous models for low-pressure ideal-gas conditions, we add the pressure-dependent compressibility factor $Z(P_g^*)$ and solubility $K_h(P_{g}^*)$ into the ideal-gas law and the Henry's law, respectively. Direct numerical simulations of the resulting PDE model are performed for a 2D rectangular domain to investigate the dynamics of convective \CO2 dissolution at large $Ra_0$.  Simple ODE models are developed to capture the mean behavior of the convecting system. To validate the PDE and ODE models, we interpret pressure-decay experiments in a PVT cell at a constant temperature of 323.15 K.

The DNS of the PDE model reveal that at high-pressure real-gas conditions, convection in closed porous media systems at large $Ra_0$ still exhibits four flow regimes characterized by the behavior of the dissolution: An early diffusion-dominated regime, a flux-growth \& plume-merging regime, a quasi-steady convective regime, and a shut-down regime, as in the ideal-gas case discussed in \citet{Wen2018JFM}.  In the quasi-steady convective regime, the dissolution flux $F$ is no longer constant due to the negative feedback of the pressure drop in the vapor phase.  However, we use the pressure-decay experiments to validate that the quantity $F/{C_s}^2$ remains constant, i.e., $F/{C_s}^2 = \alpha Ra_0$, providing a new direction for determination of the convective dissolution flux of CO$_2$ in porous media at reservoir conditions.  Moreover, the simulations indicate that in the shut-down regime the horizontal-mean concentration is well-mixed and nearly constant with depth.  Based on the above properties, nonlinear ODE models are built to predict the evolution of gas pressure and dissolution flux in the quasi-steady convective and shut-down regimes at high-pressure real-gas conditions.  Our simulations show that both the PDE and ODE models developed in this work can quantitatively capture the mean behavior of the convective \CO2 dissolution measured from experiments.  Finally, our models also reveal that for increasing gas pressure $P_{g}$ in closed systems, the negative feedback of the pressure drop is weakened due to the decrease of the solubility constant $K_h(P_g)$ at the real-gas conditions.

Assuming constant \CO2 concentration at the top boundary (or constant gas pressure in closed systems), previous 2D numerical simulations have shown $F=\alpha Ra \approx 0.017Ra_0$ for high-Rayleigh-number convective \CO2 dissolution.  Although this linear scaling has been verified in some laboratory experiments in Hele-Shaw cells or granular porous media \citep{Liang2018,DePaoli2020}, different values of the prefactor $\alpha$ are obtained since the experiments are performed using analog fluids at the atmospheric-pressure condition.  In this study our four experiments, which are performed using \CO2 and water at the reservoir conditions, show that the prefactor $\alpha$ varies from 0.012 to 0.023.  To the best of our knowledge, this is the first experimental study to evaluate the prefactor of $F\sim \alpha Ra_0$ for high-Rayleigh-number convective \CO2 dissolution at reservoir conditions.  The variation of the value of $\alpha$ in our experiments can be due to some inherent properties of the porous media, e.g., the random packing of glass beads and the effect of mechanical dispersion \citep{Liang2018,Wen2018PRFluids,Gasow2020,DePaoli2020,Hewitt2020}, etc.  Further systematic experiments are needed to show the effects of various experimental parameters (e.g., initial pressure, height of the liquid layer, and bead size, etc.) on $\alpha$.

The method presented here is currently the only way to experimentally determine the convective dissolution flux in the actual CO$_2$-brine system. This opens the door to study a variety of important processes that cannot be studied in analog systems. For example, our method could be used to develop a set of experiments to study the effect of the capillary transition zone on the convective dissolution process and validate numerical simulations \citep{Martinez2016}. Working with CO$_2$ and brine also makes it possible to study chemical effects on the convective dissolution rates that cannot be addressed by analog systems. These include the effect of impurities such as hydrogen sulfide (H$_2$S) in the gas, the composition of the brine and potentially geochemical interactions with the rock matrix.  \\

\noindent
\textbf{Acknowledgments}\\

\noindent
The work at UM was supported in part by US National Science Foundation award DMS-1813003.  The work at UT was supported as part of the Center for Frontiers in Subsurface Energy Security, an Energy Frontier Research Center funded by the U.S. Department of Energy, Office of Science, Basic Energy Sciences under Award number DE-SC0001114. The material is also based in part upon work supported by the U.S. Department of Energy National Energy Technology Laboratory under Grant Number DE-FC26-05NT42590. This project is managed and administered by the Bureau of Economic Geology, Jackson School of Geosciences, The University of Texas at Austin under a subcontract to the Southern States Energy Board.  The work at USC was supported by the Center of Geological Storage of \CO2 (GSCO2) and the Center for Mechanistic Control of Unconventional Formations (CMC-UF), an Energy Frontier Research Center funded by the U.S. Department of Energy, Office of Science.  Data sets related to this article can be found at DOI: 10.17632/r5z8578btn.1, an open-source online data repository hosted at Mendeley Data.




\bibliographystyle{elsarticle-harv} 
\bibliography{ClosedSystem}


%
%
%

\end{document}